\documentclass{PoS}

\usepackage{graphicx}
\usepackage{amsmath}

\newlength{\mywidth}
\newlength{\mywidthhalf}
\title{2+1 flavor QCD with the fixed point action in the $\epsilon$-regime}

\ShortTitle{2+1 flavor QCD in the $\epsilon$-regime}

\author{Peter~Hasenfratz$^a$, \speaker{Dieter~Hierl}$^b$, Vidushi~Maillart$^a$, Ferenc~Niedermayer$^a$, Andreas~Sch\"afer$^b$, Christof~Weiermann$^a$, Manuel~Weingart$^a$\\
BGR (Bern-Graz-Regensburg) Collaboration\\\\
\llap{$^a$} Institute for Theoretical Physics, University of Bern, CH-3012 Bern, Switzerland\\
\llap{$^b$} Institute for Theoretical Physics, University of Regensburg, D-93040 Regensburg, Germany\\
E-mail: \email{dieter.hierl@physik.uni-regensburg.de}}

\abstract{We generated configurations with the approximate fixed-point Dirac operator $D_\mathrm{FP}$ on a $12^4$ lattice with $a \approx 0.13\;$fm where the scale was set by $r_0$. The distributions of the low lying eigenvalues in different topological sectors were compared with those of the Random Matrix Theory which leads to a prediction of the chiral condensate.}

\FullConference{
The XXV International Symposium on Lattice Field Theory\\
July 30 - August 4 2007\\
Regensburg, Germany}

\begin{document}

\section{Introduction}

The $\epsilon$-expansion, where $m_\pi L \ll 1$ and $4\pi F^2 L^2 \gg 1$, is determined by Goldstone boson physics. For that reason it is an excellent tool to determine the low-energy constants of ChPT of QCD. In order to reach this regime the simulation should be done with sufficiently small quark masses. This requires a Dirac operator with good chiral behavior at the actual lattice spacing. This became possible in full QCD only recently \cite{DeGrand:2006nv,DeGrand:2007mi,Fukaya:2007yv,Hasenfratz:2007yj}.

We consider $2+1$ light flavor QCD applying the parametrized fixed-point (FP) action \cite{Hasenfratz:2007yj}. The exact FP action has nice features, most importantly, it has exact chiral symmetry. The parametrized FP action is an approximation which gave very promising results in the quenched approximation, in particular good scaling even at $a = 0.15\;$fm and the spectrum of the FP Dirac operator was close to that required by chiral symmetry.

The FP Dirac operator satisfies the Ginsparg-Wilson relation:
\begin{equation}
D \gamma_5 + \gamma_5 D = D \gamma_5 2 R D \;,\label{eq:ginwil}
\end{equation}
where $R$ is a local operator living on the hypercube. The quark mass is introduced as usual for Ginsparg-Wilson operators:
\begin{equation} \label{eq:Dm}
D(m) = D + m \left( \frac{1}{2R} - \frac{1}{2} D \right)\;.
\end{equation}

The parametrized fixed-point gauge action and Dirac operator $D_\mathrm{FP}$ involve a special smearing with projection to $\mathrm{SU}(3)$. Therefore we cannot use a hybrid Monte Carlo algorithm. We use a partially global update procedure with nested accept/reject steps \cite{Hasenfratz:2005tt} relying on algorithmic developments in \cite{Hasenbusch:2001ne, de Forcrand:1998sv, Hasenfratz:2002jn, Kennedy:1985pg, Kennedy:1988yy, Joo:2001bz, Montvay:2005tj}. The contribution of the $\sim 100$ lowest lying modes to be determined is calculated exactly, the rest is treated stochastically. As a spinoff, for all the configurations in the Markov chain we have the low-lying eigenvectors which can be used in the analysis.

\subsection{Remarks on the Markov chain and numerical implementation}

In the algorithm the trial configurations offered in the last stochastic accept/reject step differ by a full Metropolis sweep and are accepted with $P_{acc} \approx 0.6$. We spent 13 minutes for one full accept/reject step in the Markov chain using 288 CPUs on the Altix in Munich. On this architecture it is very important to get into the cache of the single cores. Therefore we had to rewrite the whole parallelization of the existing code already running on diverse other machines while implementing it for the Altix. We use the BiCGStab($l$) inverter in the stochastic estimator where $l$ is tuned depending on the last number of inversion steps.

At the moment we have $\sim 4000$ configurations in the Markov chain using the partially global update algorithm \cite{Hasenfratz:2005tt}. In order to reduce the autocorrelation time (and the expenses of the analysis) we only take every tenth configuration minimizing the autocorrelation length and get $\sim 400$ pruned configurations.

\subsection{Technicalities - $12^4$ lattice}

\begin{figure}[ht]
\begin{center}
\includegraphics[width=\mywidth]{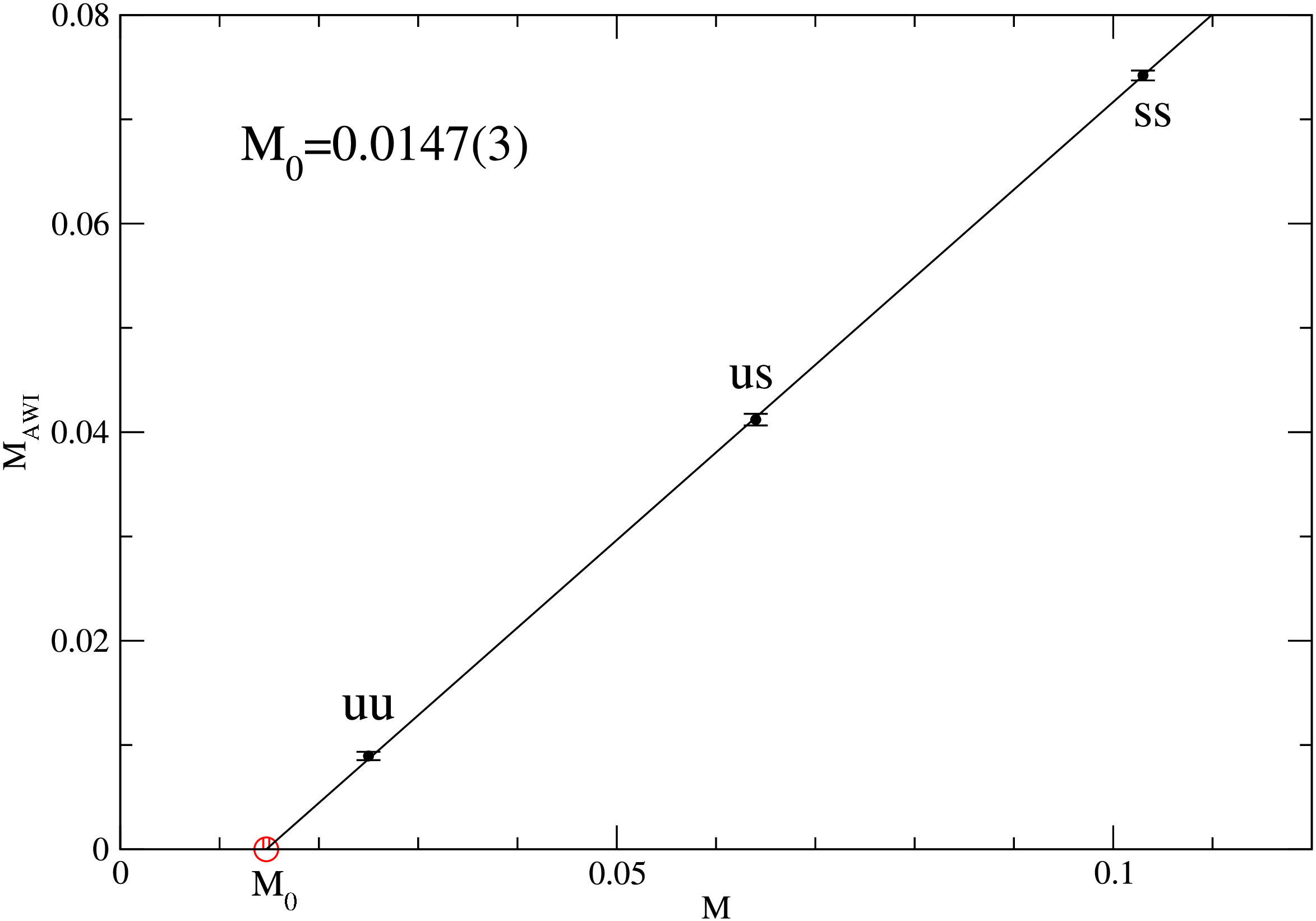}
\end{center}
\caption{{}The AWI mass vs. the average quark mass $M = (m_f+m_{f'})/2$.\label{fig:awi}}
\end{figure}

Our lattice spacing is $a=0.129(5)\;$fm determined from the Sommer parameter $r_0=0.49\;$fm. Therefore our $12^4$ lattice has a volume of $V \approx (1.55\;\mathrm{fm})^4$. We calculated the additive mass shift of $aM_0=0.0147(3)$ via the axial Ward identity:
\begin{equation} \label{eq:Ward}
\partial_t \langle A_{ff'}(t)P_{ff'}(0) \rangle = 2 M_\mathrm{AWI} \langle P_{ff'}(t)P_{ff'}(0) \rangle\;,\qquad \mbox{where } ff'=uu\,, us\,, ss\;.
\end{equation}
In Fig.~\ref{fig:awi} we plotted the AWI mass as a function of the average quark mass $M = (m_f + m_{f'})/2$, together with a linear fit. The intercept with the horizontal axis gives the additive mass renormalization of the quark masses. One would reach the chiral limit at this value of the mass parameter in Eq.~\eqref{eq:Dm}. The points $uu$, $us$ and $ss$ refer to the corresponding flavors in Eq.~\eqref{eq:Ward}. Subtracting the mass shift $M_0$ we get for the bare masses $m_{ud} = 16\;$MeV and $m_{s} = 137\;$MeV.

\section{Eigenvalues of the Dirac operator}

\begin{figure}[ht]
\begin{center}
\includegraphics[width=0.7\mywidth]{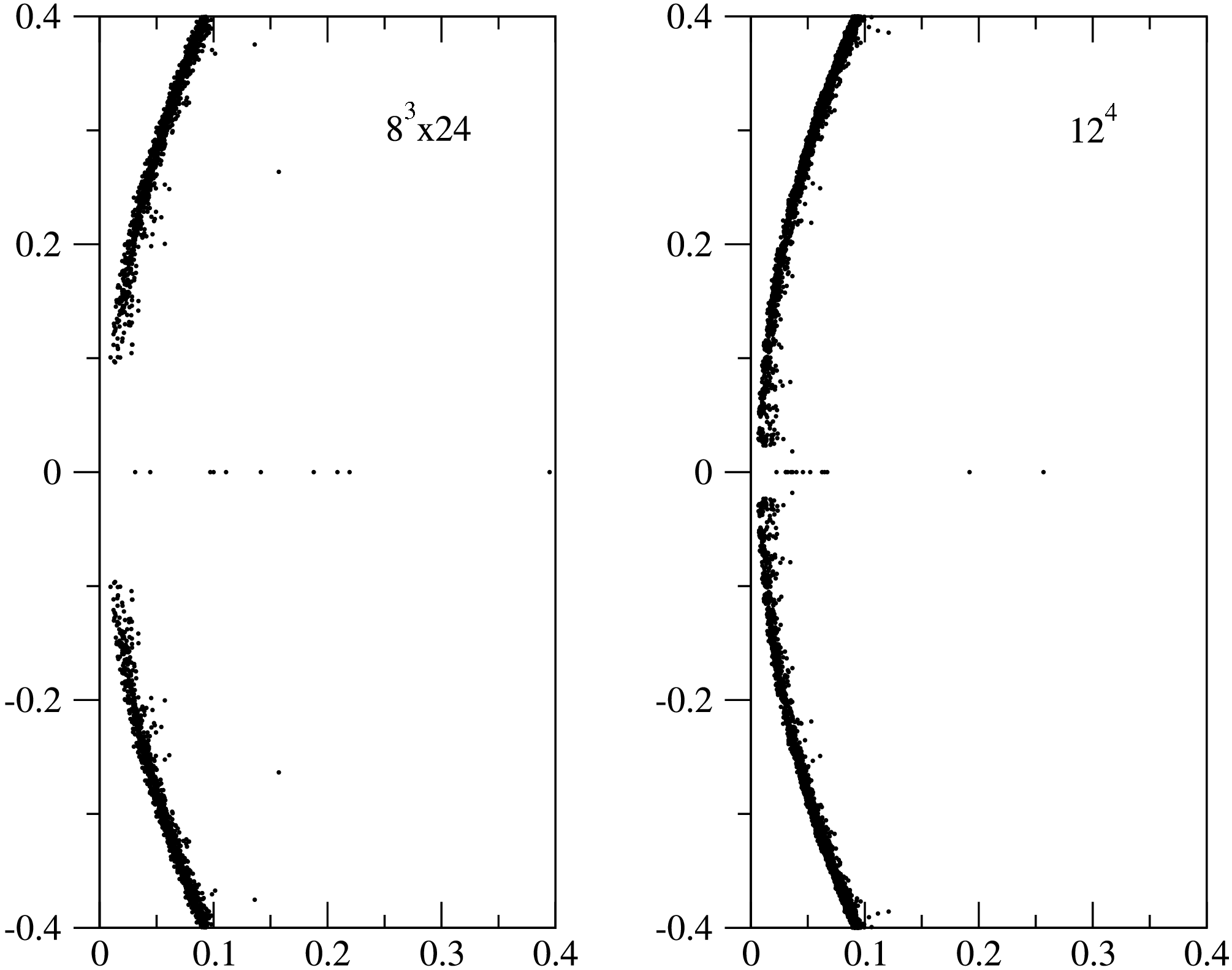}
\end{center}
\caption{{}The left plot shows the low-lying eigenvalues of the FP Dirac operator on 50 configurations at the volume $8^3 \times 24$ and the plot on the right hand side displays the same for $12^4$ configurations.\label{fig:spectrum}}
\end{figure}

The FP Dirac operator follows Eq.~\eqref{eq:ginwil}. On low-lying eigenvectors the operator $2R$ is close to $1$ and we ignore it in this presentation for simplicity, i.e.~we set $2R = 1$ in our discussion. Then the low-lying complex spectrum of $D_\mathrm{FP}$ lies on a circle to a good approximation, while some of the real eigenvalues are scattered away from the origin.

\begin{figure}[ht]
\begin{center}
\includegraphics[angle=-90,width=\mywidth]{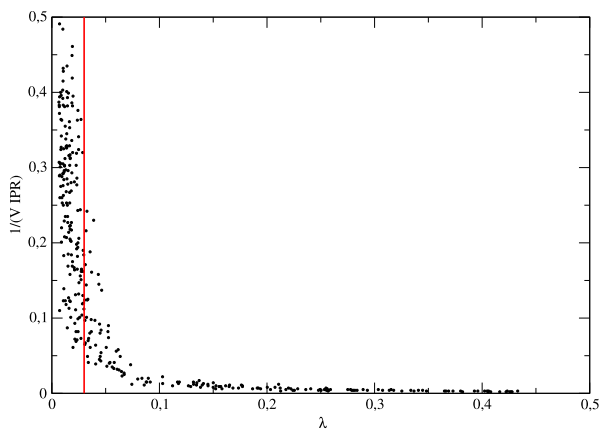}
\end{center}
\caption{{}The relative effective volume occupied by the wave function ($(V \times \mathrm{IPR})^{-1}$) vs. the eigenvalue of $D_\mathrm{FP}$ for real modes.\label{fig:IPR}}
\end{figure}

It is expected that the size of a effective support of the wave function and the position of the real eigenvalue are strongly correlated. Thus we introduce the inverse participation ratio (IPR) by
\begin{equation}
IPR = \sum_x ||\psi^{(\lambda)}(x)||^4 \,,\qquad \mbox{where} \quad ||\psi^{(\lambda)}(x)||^2 = \sum_{i=1}^{12} |\psi_i^{(\lambda)}(x)|^2
\end{equation}
for a normalized eigenvector $\psi^{(\lambda)}$. We plot the participation ratio $p = {(V \times IPR)}^{-1}$, where we get $p \to \mathcal{O}(1)$ for delocalized and $p \to \mathcal{O}(1/V)$ for localized eigenvectors $\psi^{(\lambda)}$.

In the analysis we used a cut $\lambda = 0.03$ (solid line in Fig.~\ref{fig:IPR}) in the sense that only those real eigenvalues which were below this value were considered to represent the topological charge $\nu$. With this cut we found $328$, $51$ and $35$ configurations for the $\nu = 0, 1, 2$ sectors, respectively. This cut is sufficiently smaller than the typical gap in the complex eigenvalues (cf.~Fig.~\ref{fig:spectrum}) and corresponds to a small participation ratio (cf.~Fig.~\ref{fig:IPR}).

\section{Comparison to RMT}

\begin{figure}[ht]
\begin{center}
\includegraphics[width=\mywidthhalf]{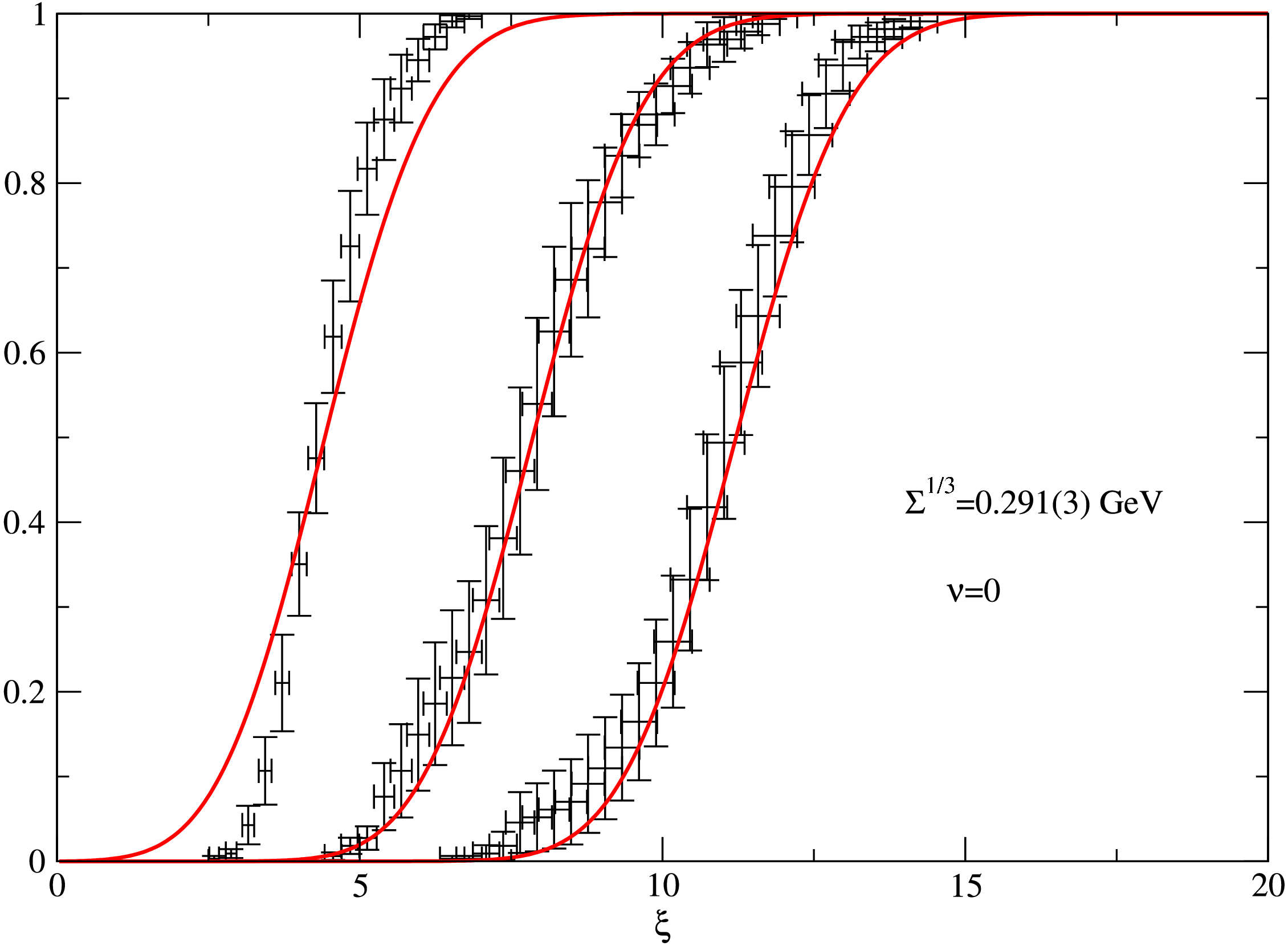}
\hfill
\includegraphics[width=\mywidthhalf]{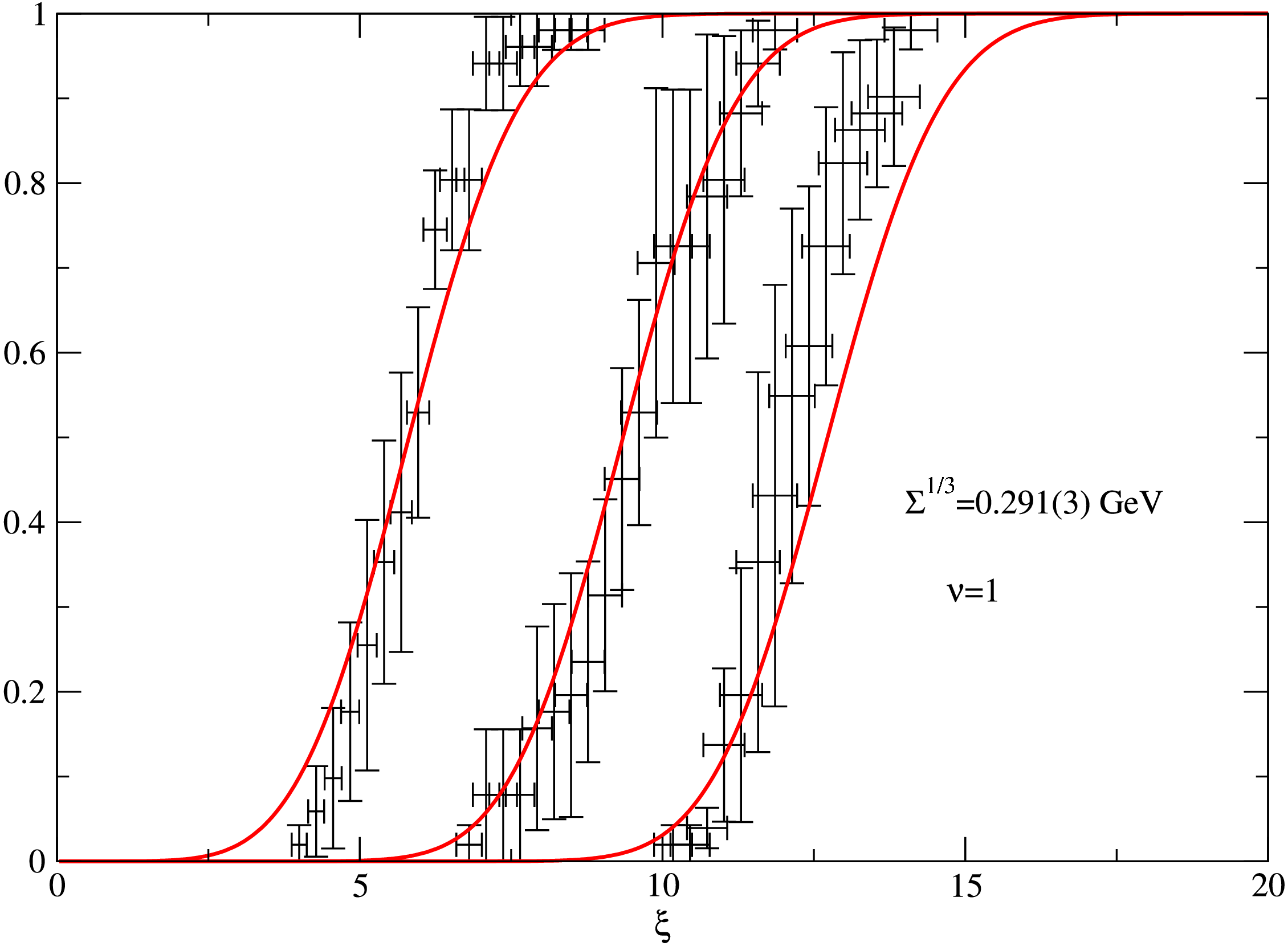}
\hfill
\includegraphics[width=\mywidthhalf]{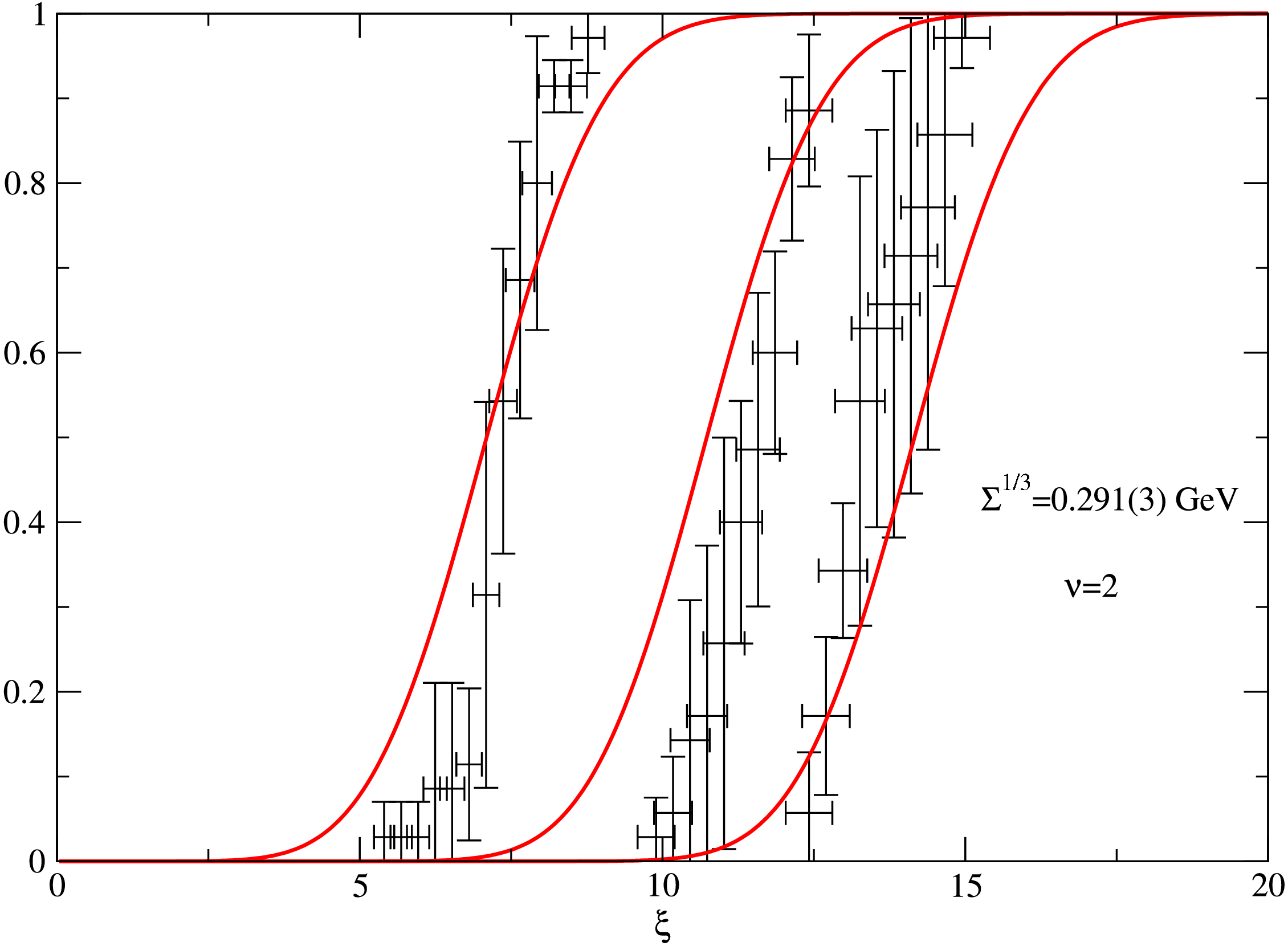}
\end{center}
\caption{{}Cumulative distributions of $\xi_{\nu k}$ in the topological sector $\nu=0,1,2$. We show the $k$-th smallest eigenvalues for $k = 1, 2, 3$.\label{fig:RMT}}
\end{figure}

We use the stereographic projection of the complex eigenvalues to the imaginary axis
\begin{equation}
i \alpha = \frac{\lambda}{1-\lambda/2}\;.
\end{equation}
We denote the $k$-th eigenvalue in the $Q_\mathrm{top}=\nu$ sector as $\alpha_{\nu k}$. Random matrix theory (RMT) predicts the distribution of the scaled eigenvalues $\xi_{\nu k}(\mu_i) = \alpha_{\nu k} \Sigma V$ depending on $\mu_i= m_i \Sigma V$, $i=1,\ldots,N_f$. The cumulative distributions of $\xi_{\nu k}$ have only one matching parameter, which is the bare condensate $\Sigma$. Fitting the distribution of the 3 lowest lying eigenvalues in the $\nu = 0, 1$ topological sector to the RMT predictions we get  $\Sigma^{1/3} = 0.286(3)(9)\;\mathrm{GeV}$.

As Fig.~\ref{fig:RMT} shows the distributions for different $\nu,k$ values are consistent with each other. Note, however, that the shape of the $\nu=0$, $k=1$ distribution is different from that of the RMT. The deviation in the shape could be a finite-size effect (which shows up at smallest eigenvalue, i.e. for largest wave-length), but this needs further investigations.

\begin{figure}[ht]
\begin{center}
\includegraphics[width=\mywidth]{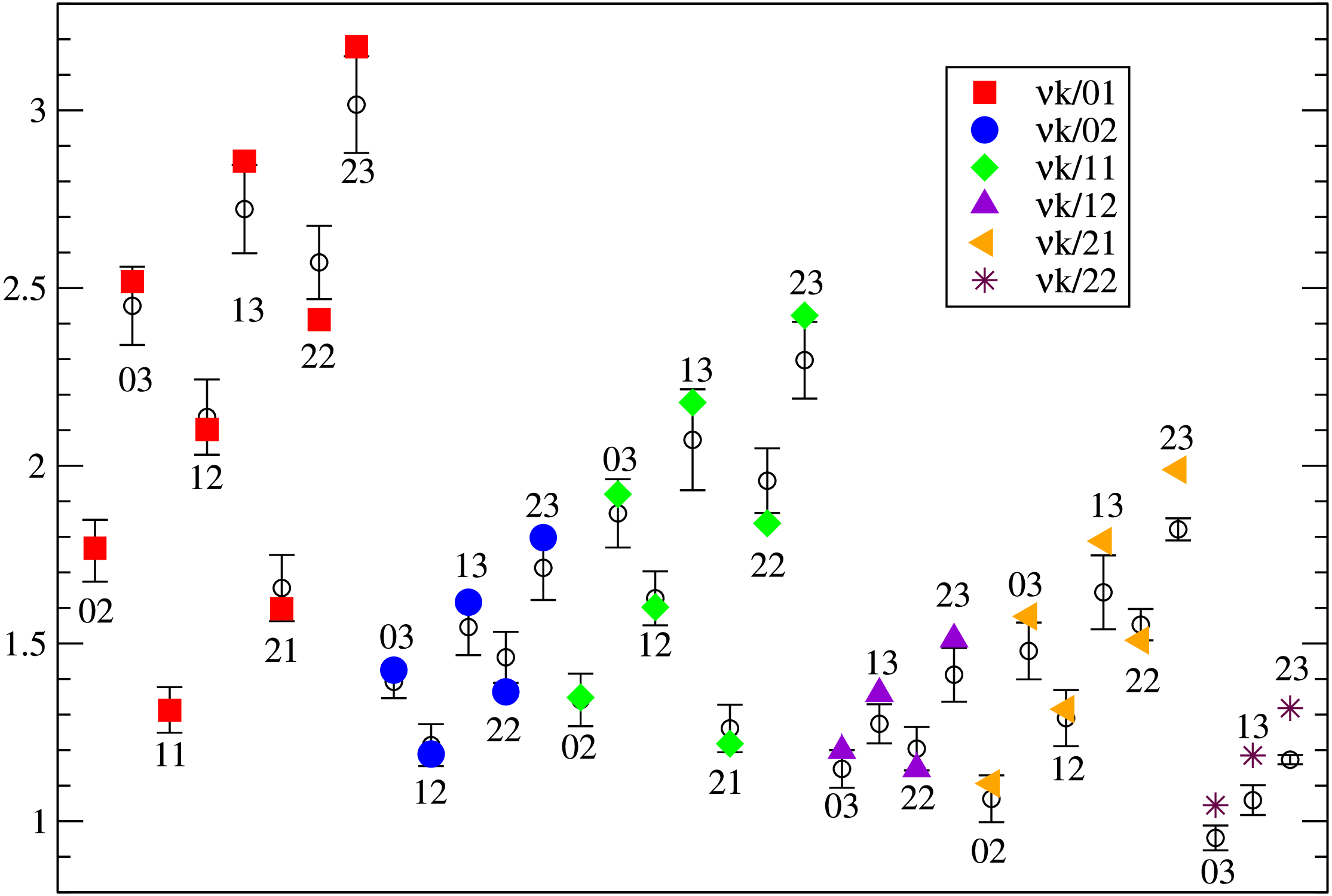}
\end{center}
\caption{{}We calculate the ratios between different averaged eigenvalues $\langle \xi_{\nu k}\rangle$ which are denoted as $\nu k$. The different symbols denote the RMT predictions for our corresponding results.\label{fig:ratio}}
\end{figure}

We also compute the ratios $\langle \xi_{\nu k}\rangle/\langle \xi_{\nu' k'} \rangle$ of the mean values and compare them to the RMT predictions. We find a generally good agreement (cf.~Fig.~\ref{fig:ratio})  even in higher topological sectors.

\subsection{Finite Volume corrections}

The determined $\Sigma$ is the condensate of QCD in the finite volume. Thus, we have to correct the finite size effects using chiral perturbation theory with $N_f=2$, assuming that in this correction the s-quark contribution is negligible \cite{Hasenfratz:2007yj, Hasenfratz:1989pk}
\begin{equation}
\Sigma = \Sigma_\infty \left( 1 + \frac{3}{2} \frac{\beta_1}{F^2 L^2} + \ldots \right)\;,
\end{equation}
where $\beta_1 = 0.14046$ is a shape coefficient. After the finite size correction we get for the infinite-volume bare condensate
\begin{equation}
\Sigma_\infty^{1/3} = 0.255(3)(9)\;\mathrm{GeV}\;.
\end{equation}

\section{Summary and Outlook}

Simulating the parametrized FP action with partially global update procedure we were able to reach sufficiently small quark masses to study the $\epsilon$-regime in the 2+1 flavor QCD. The distribution of the low-lying eigenvalues in different topological sectors is in good agreement with the RMT predictions, although the identification of the topological charge is somewhat ambiguous.

The next steps are to compute the low energy constant $F$ from the $\langle PP \rangle$ correlator, to calculate the corresponding $Z$ factors and to complete a lattice simulation on a $12^3 \times 24$ lattice with comparable lattice spacing in the $\delta$-regime.

\section{Acknowledgment}

This work was supported in part by Schweizerischer Nationalfonds and by DFG (FG-465). We acknowledge the support and computing resources at CSCS, Manno and LRZ (project h0323a).


\begin{thebibliography}{99}

\bibitem{DeGrand:2006nv}
T.~DeGrand, Z.~Liu and S.~Schaefer,
Phys.\ Rev.\ D {\bf 74} (2006) 094504
[Erratum-ibid.\  D {\bf 74} (2006) 099904]
[arXiv:hep-lat/0608019].

\bibitem{DeGrand:2007mi}
T.~DeGrand and S.~Schaefer,
arXiv:0709.2889 [hep-lat].

\bibitem{Fukaya:2007yv}
H.~Fukaya {\it et al.},
Phys.\ Rev.\  D {\bf 76} (2007) 054503
[arXiv:0705.3322 [hep-lat]].

\bibitem{Hasenfratz:2007yj}
P.~Hasenfratz, D.~Hierl, V.~Maillart, F.~Niedermayer, A.~Schafer,
C.~Weiermann and M.~Weingart, 
arXiv:0707.0071 [hep-lat].

\bibitem{Hasenfratz:2005tt}
A.~Hasenfratz, P.~Hasenfratz and F.~Niedermayer,
Phys.\ Rev.\  D {\bf 72}, 114508 (2005)
[arXiv:hep-lat/0506024].

\bibitem{Hasenbusch:2001ne}
M.~Hasenbusch,
Phys.\ Lett.\  B {\bf 519}, 177 (2001)
[arXiv:hep-lat/0107019].

\bibitem{de Forcrand:1998sv}
P.~de Forcrand,
Nucl.\ Phys.\ Proc.\ Suppl.\  {\bf 73}, 822 (1999)
[arXiv:hep-lat/9809145].

\bibitem{Hasenfratz:2002jn}
A.~Hasenfratz and F.~Knechtli,
Comput.\ Phys.\ Commun.\  {\bf 148}, 81 (2002)
[arXiv:hep-lat/0203010].

\bibitem{Kennedy:1985pg}
A.~D.~Kennedy and J.~Kuti,
Phys.\ Rev.\ Lett.\  {\bf 54}, 2473 (1985).

\bibitem{Kennedy:1988yy}
A.~D.~Kennedy, J.~Kuti, S.~Meyer and B.~J.~Pendleton,
Phys.\ Rev.\  D {\bf 38}, 627 (1988).

\bibitem{Joo:2001bz}
B.~Joo, I.~Horvath and K.~F.~Liu,
Phys.\ Rev.\  D {\bf 67}, 074505 (2003)
[arXiv:hep-lat/0112033].

\bibitem{Montvay:2005tj}
I.~Montvay and E.~Scholz,
Phys.\ Lett.\  B {\bf 623}, 73 (2005)
[arXiv:hep-lat/0506006].

\bibitem{Hasenfratz:1989pk}
P.~Hasenfratz and H.~Leutwyler,
Nucl.\ Phys.\  B {\bf 343}, 241 (1990).

\end{thebibliography}
\end{document}